\documentclass{article}
\usepackage{spconf,amsmath,graphicx,hyperref}
\usepackage{cite}

\usepackage{amsmath,amssymb,bm,amsthm,mathrsfs,dsfont}
\usepackage{tikz}
\usepackage{algorithmicx}
\usepackage{algpseudocode}
\usepackage[norelsize, linesnumbered, ruled, vlined, lined, boxed]{algorithm2e}

\allowdisplaybreaks
\usepackage[nodisplayskipstretch]{setspace} 
\title{Deep Reinforcement Learning for  Dynamic Sensing and Communications 
}
\name{
Abolfazl Zakeri$^{1}$,
Nhan Thanh Nguyen$^{1}$,
Ahmed Alkhateeb$^{2}$,
and Markku Juntti$^{1}$
}

\address{
$^{1}$ Centre for Wireless Communications – Radio Technologies (CWC-RT), University of Oulu, Finland \\
Email: \{abolfazl.zakeri, nhan.nguyen, markku.juntti\}@oulu.fi \\
$^{2}$ School of Electrical, Computer, and Energy Engineering, Arizona State University, USA \\
Email: alkhateeb@asu.edu
}

\begin{document}
%
\maketitle
\begin{abstract}
Environmental sensing can significantly enhance mmWave communications by assisting beam training, yet its benefits must be balanced against the associated sensing costs. To this end, we propose a unified machine learning framework that dynamically determines when to sense and leverages sensory data for beam prediction. Specifically, we formulate a joint sensing and beamforming problem that maximizes the average signal-to-noise ratio under an average sensing budget. Lyapunov optimization is employed to enforce the sensing constraint, while a deep Q-Network determines the sensing slots. A pretrained deep neural network then maps the sensing data to optimal beams in the codebook. Simulations based on the real-world DeepSense dataset demonstrate that the proposed approach substantially reduces sensing overhead while maintaining satisfactory communications performance.
\end{abstract}
\section{Introduction}\label{sec:intro}\vspace{-0.25cm}
Beam training is essential for reliable performance in wireless systems, particularly at millimeter-wave (mmWave) frequencies, where links are highly sensitive to path loss and blockages.  
Recently, multimodal sensing-aided communications, which leverage sensory data such as visual, LiDAR, and radar measurements, have attracted growing interest \cite{Ahmed_deepsense, R_Heath_mag, multimodality_mmwave_mag}.  
By enhancing environmental perception and situational awareness, multimodal sensing helps reduce beam training overhead \cite{RHeath_BT_multiuser, Ahmed_seman_dis} and improve beam alignment \cite{multimodality_connected_vechile}. These benefits are especially valuable in high-mobility settings where proactive line-of-sight (LoS) prediction and beam selection are critical for sustaining reliable connectivity.
\par
Recent studies in multimodal sensing-aided communications \cite{Ahmed_vision, multimodality_connected_vechile, latin_amr_RH, transformer_sensing, salehihikouei2024leveraging, zakeri_globe25sensing, Debbah_25_llm, Ahmed_LiDar_COML, DL_vtc24, Khaled_25, RHeath_BT_multiuser, Walid_ML25} have primarily focused on  beam prediction.  
Patel \textit{et al.} \cite{RHeath_BT_multiuser} showed that sensor-aided deep learning enables efficient channel estimation for multi-user mmWave systems, achieving interference-free beamforming and improved spectral efficiency.  
Jiang \textit{et al.} \cite{Ahmed_LiDar_COML} demonstrated that LiDAR-aided machine learning can predict and track beams in real vehicular settings with low training overhead, a result later extended to multimodal prediction \cite{Ahmed_vision, DL_vtc24}.  
These works show that multimodal sensing not only facilitates beam training  but also improves beamforming performance.
\\\indent
Nonetheless, most prior works overlook the optimization of the sensing process itself, often assuming that sensory data is continuously available.  
While such assumptions yield useful insights into how different modalities can aid beam training, they leave open a fundamental question:  
\textit{When should sensing be performed, and which modality should be employed to efficiently sustain communication performance?} 
Addressing this question requires explicitly modeling the tradeoff between the sensing rate (i.e., how often to sense) and the sensing modality (i.e., which modality to employ) versus the resulting communications performance. This, in turn, necessitates algorithms that support \textit{dynamic and adaptive joint} sensing and communications design.
\\\indent
In our previous work~\cite{zakeri_globe25sensing}, we introduced a constrained sensing and dynamic beamforming approach.~That approach requires an exhaustive search for sensing decisions at each slot and relies solely on position-based channel modeling. To overcome these limitations, in this work, we propose a two-module learning framework: a Deep Q-Network (DQN) for sensing decisions and a deep neural network (DNN) for beam prediction (see Fig.~\ref{fig_proposed_method}).  
The objective is to dynamically determine both when to sense and which beam to select to maximize average signal-to-noise-ratio (SNR), subject to an average sensing constraint.  
Specifically, we first pretrain the DNN on available data with optimal beam labels, then apply Lyapunov queue stability to enforce the sensing constraint, thereby transforming the problem into a Markov decision process solved by the DQN in conjunction with the pretrained DNN.  
Preliminary results on the DeepSense position dataset show that our framework can reduce sensing frequency by nearly 50\% while maintaining beam prediction accuracy comparable to always sensing in every slot.
\vspace{-1 em}
\section{System Model and Problem Formulation}\label{system_model}\vspace{-0.25cm}
We consider a downlink mmWave communications system including a base station (BS) and a single-antenna mobile user equipment (UE). The BS is equipped with $N$ antennas and a sensing unit. At time slot ${t=1,2,\dots}$, the BS employs analog beamforming vector $\mathbf{w}(t) \in \mathcal{W}$ for signal transmission, where $\mathcal{W} = \{\mathbf{w}_1, \dots, \mathbf{w}_M\}$ denotes the codebook of $M$ candidate beamforming vectors with $\|\mathbf{w}_m\|_{2}^{2} = 1, \forall\, m$.
\\\indent 
Denote by $\mathbf{h}(t)\in\mathbb{C}^{N\times 1}$ the channel between the BS and the UE at time slot $t$. Furthermore, let ${s}(t)\in\mathbb{C}$ be the transmit data symbol from the BS to the UE, $\mathbb{E}\{|{{s}(t)}|^2\}=P$ with $P$ being the transmit power.
The received signal at the UE is then given by 
\begin{align}
    y(t)= \mathbf{h}^{\textsf{H}}(t) \mathbf{w}(t){s}(t) + n(t),
\end{align}
where  $n(t)\in \Bbb{C}$ is additive white Gaussian noise (AWGN) following the distribution $\mathcal{CN}(0,\sigma^2)$, with $\sigma^2$ denoting the noise variance at the UE's receiver.
\begin{figure}[t!]
 \centering
  \includegraphics[width=7.0cm]{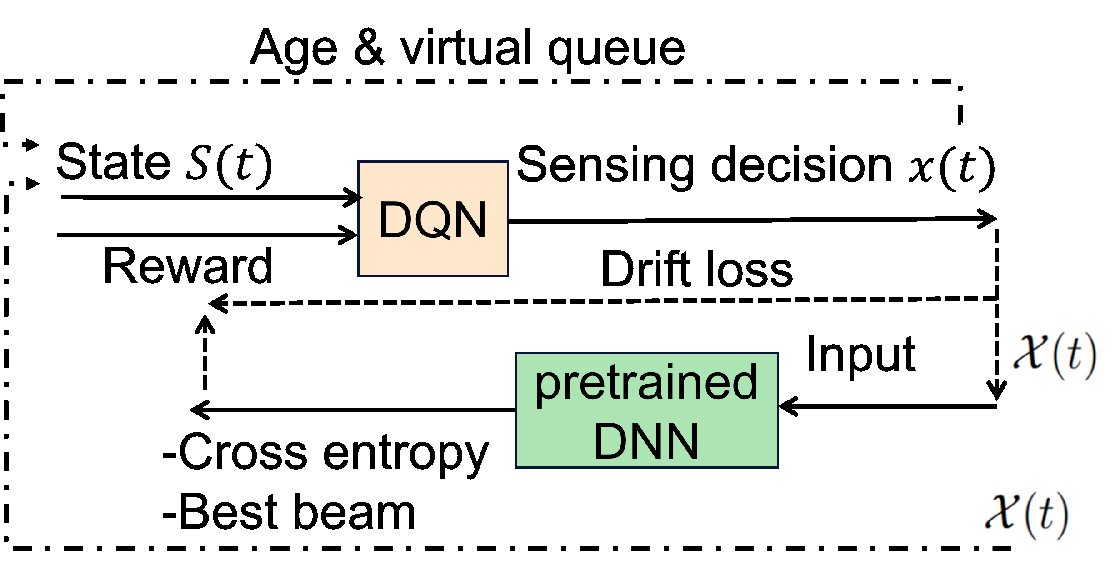}
  \caption{A schematic of the proposed two-module machine learning method for dynamic sensing and beam prediction}
  \label{fig_proposed_method}
\end{figure}
\\\indent
\textit{Problem Formulation:}  
The primary task in multimodal sensing for communications is to exploit sensory data (e.g., position, images, LiDAR) to select the best beam from $M$ candidates that maximizes the UE’s received SNR, i.e., $|\mathbf{h}^{\mathsf{H}}(t)\mathbf{w}(t)|^2$.  
This is typically done by training a DNN on collected sensory data with corresponding best-beam labels.  
In this paper, we extend this task to a dynamic \emph{joint sensing and beam prediction} problem.
\\\indent
Let $x(t)\in\{0,1\}$ denote the sensing decision at slot $t$, where $x(t)=1$ indicates that sensing is performed and $x(t)=0$ otherwise.  
Performing sensing provides fresh data for beam prediction but incurs a cost $c$, which accounts for, e.g., power consumption and processing overhead of data acquisition.  
Since using recent data generally improves beam prediction accuracy, this naturally leads to a tradeoff between sensing frequency (i.e., data acquisition rate) and communication performance.  
We formulate this tradeoff as the following optimization problem:
\vspace{- 1em}
\begin{subequations}
\label{op_1}
\begin{align}
\underset{\{m(t), x(t)\}_{t=1,2,\ldots}}{\mbox{maximize}}~~~ & 
\limsup_{T\rightarrow \infty } \frac{1}{T} \sum_{t=1}^{T} 
\mathbb{E}\!\left[|\mathbf{h}^{\mathsf{H}}(t)\mathbf{w}_{m(t)}|^2\right] \label{eq_obj_fun} \\
\mbox{subject to}~~~~~~~~ &  
\limsup_{T\rightarrow \infty } \frac{1}{T} \sum_{t=1}^{T} 
\mathbb{E}\!\left[cx(t)\right] \leq c^{\mathsf{max}}, \label{eq_cons_sensing}
\end{align}
\end{subequations}
where $m(t)\in\{1,\dots,M\}$ is the selected beam index, $x(t)\in\{0,1\}$ is the sensing decision, and $c^{\mathsf{max}}$ denotes the sensing cost budget.  
The implicit effect of $x(t)$ lies in determining whether \textit{fresh} sensory data is available for beam selection.
\\\indent 
The main challenge in solving problem~\eqref{op_1} is that the instantaneous channel information $\mathbf{h}(t)$ is not available.  
As an alternative, multimodal sensory data combined with machine learning (ML) can be leveraged to learn the best beam selections.  
To this end, we propose a two-module ML framework (see Fig.~\ref{fig_proposed_method}): a DQN for sensing decisions and a DNN for beam prediction, as detailed in the next section.

\section{Proposed Joint Sensing and Beam Prediction Approach }

As illustrated in Fig.~\ref{fig_proposed_method}, our framework consists of two ML modules:  
(i) a DQN that learns over time the optimal sensing actions, and  
(ii) a DNN that predicts the best beam given the available sensory data.  
The DNN is first trained offline using sensory data samples from all time slots.  
Then, the DQN is trained with the pretrained DNN as its beam prediction module.  
It is important to note that the DQN’s sensing decisions must satisfy the sensing cost constraint in \eqref{eq_cons_sensing}.

To enforce the average constraint~\eqref{eq_cons_sensing}, we adopt the Lyapunov queue stability framework~\cite[Ch.~2]{Neely_Sch}.  
Let $Q(t)$ denote the virtual queue associated with constraint~\eqref{eq_cons_sensing} at slot $t$, whose evolution is given by 
\begin{equation}\label{eq_virtualQueue}
    Q(t+1) = \max \Big[ Q(t) + x(t) - \frac{c^{\mathsf{max}}}{c},\, 0 \Big].
\end{equation}
Here, $Q(t)$ evolves as a queue with arrival rate $x(t)$ and service rate $\tfrac{c^{\mathsf{max}}}{c}$.  
According to~\cite[Ch.~2]{Neely_Sch}, the time-average constraint~\eqref{eq_cons_sensing} is satisfied if the queue is \emph{strongly stable}, i.e., 
$
\limsup_{T\rightarrow \infty} \frac{1}{T}\sum_{t=1}^T \mathbb{E}\{Q(t)\} < \infty.
$

We now define the Lyapunov function and its drift to facilitate the stability analysis.
Let $L(Q(t))=\tfrac{1}{2}Q^2(t)$ denote the quadratic Lyapunov function~\cite[Ch.~3]{Neely_Sch}.  
Stabilization of the virtual queue can be achieved by minimizing the expected change of this function across slots~\cite[Ch.~3]{Neely_Sch}.  
The one-slot conditional Lyapunov drift, denoted by $\Delta(t)$, is defined as the expected change in the Lyapunov function over one slot, conditioned on the current system state $S(t)$, i.e.,
\begin{equation}\label{eq_drift}
    \Delta(t) \triangleq \mathbb{E}\!\left[ L(Q(t+1)) - L(Q(t)) \,\big|\, S(t) \right].
\end{equation}

Having the drift is defined, next, we explain the cost function used for the training of the DQN module in Fig. \ref{fig_proposed_method}. We have used a similar idea in \cite{Zakeri_Journal_Relay} and apply a drift-plus-penalty notion \cite{Neely_Sch} to define the cost function for each state $S(t)$ and action $x(t)$, denoted by $C(S(t),x(t))$. Thus, the objective of DQN can be expressed as 
\begin{align}
J(\pi)\;=\;\mathbb{E}\!\left[\sum_{t=0}^{\infty}\gamma^t\,C(S(t),x(t))\right],
\qquad 0\le \gamma<1. 
\end{align}
where $\pi$ denotes the policy, i.e., a mapping from each state $S(t)$ to the binary sensing action $x(t)$, determined by the DQN via the $\arg\max$ operator on its output layer. Moreover, $\gamma$ is the discount factor. The expectation is taken over the random trajectories induced by policy $\pi$, due to (possibly) stochastic action selection and environment dynamics.
\textit{DQN Training Cost Function:}  
The immediate cost function for DQN, denoted as $C(S(t),x(t))$, consists of two components:  
(1) the \textit{beam prediction loss} associated with the DNN module, since the DQN’s sensing decision determines the DNN input and thereby affects beam prediction accuracy; and  
(2) the penalty associated with the average sensing constraint \eqref{eq_cons_sensing}, captured by the drift term $L(Q(t+1)) - L(Q(t))$.  For the beam prediction loss, let the softmax output of the DNN be given by $\Pr\{m^*(t)=i\}=p_i(t)$, where
\begin{align}
    p_i(t) = \mathsf{softmax}(z_i(t)) 
    \triangleq \frac{e^{z_i(t)}}{\sum_{j=1}^M e^{z_j(t)}}, i=1,\dots,M,
\end{align}
and $z_i(t)$ is the $i$-th logit (i.e., raw output of the model). Ideally, we desire $p_i(t)$ to be a delta function, i.e., ${p_i(t) = \delta(i - m^*(t))}$, where $m^*(t)$ is the optimal beam index at time $t$.  
Accordingly, the beam prediction loss is defined using cross-entropy:
\begin{align}\label{eq_entropyloss}
    \mathcal{L}(t) \triangleq -\sum_{i=1}^M \delta(i- m^*(t)) \log(p_i(t)) 
    = -\log(p_{m^*(t)}(t)).
\end{align}
The immediate cost function is then given by
\begin{align}\label{eq_dqn_cost}
    C(S(t),x(t)) \triangleq V \mathcal{L}(t) + \big( L(Q(t+1)) - L(Q(t)) \big),
\end{align}
where $V$ is a non-negative parameter chosen to desirably adjust a trade-off between the size of the virtual queue and the beam prediction accuracy.

\textit{DQN State $S(t)$:}  
Let $\mathcal{X}(t)$ denote the \textit{input} to the DNN module in Fig.~\ref{fig_proposed_method}, e.g., a 2D position concatenated with RGB image features.  
Let $\mathcal{X}_{\mathsf{curr}}(t)$ denote the \textit{current} data at slot $t$ (e.g., current position and RGB image). Then,
\begin{align}\label{eq_ml_input}
   \mathcal{X}(t) = 
   \begin{cases}
     \mathcal{X}_{\mathsf{curr}}(t), & \text{if } x(t)=1, \\
     \mathcal{X}_{\mathsf{old}}(t), & \text{if } x(t)=0,
   \end{cases}
\end{align}
where $\mathcal{X}_{\mathsf{old}}(t) = \mathcal{X}_{\mathsf{curr}}(t')$ for the latest $t'$ such that $x(t')=1$.  The DQN state is then defined as
\begin{align}
    S(t) \triangleq \big( \mathcal{X}(t),\, Q(t),\, \theta(t) \big),
\end{align}
where $\theta(t)$ is the \textit{age of the most recent sample}, evolving as
\begin{align}\label{eq_agedynamic}
   \theta(t+1) = 
   \begin{cases}
     1, & \text{if } x(t)=1, \\
     \theta(t)+1, & \text{if } x(t)=0.
   \end{cases}
\end{align}
\indent In Section~\ref{sec_numres}, we numerically evaluate the impact of including the age of information $\theta(t)$ in the DQN state (see Fig.~\ref{fig_avrgaccu_vs_alpha}).  
The implementation details of both the DQN and DNN modules are provided in the next section.
\\\indent 
We summarize our method in Alg.~\ref{alg:dnn_dqn_training}. After initialization of system parameters and neural network models, loading a dataset, two main steps are performed. In Step~(1), the DNN is trained in a supervised manner using labeled beam data, where mini-batches are sampled, predictions are generated, and parameters are updated via cross-entropy backpropagation. In Step (2), the DQN is trained: at each time step, an action is selected using the $\epsilon$-greedy strategy, i.e., choosing a random action with probability $\epsilon$ for exploration or the best action with probability $1-\epsilon$ for exploitation. The chosen action updates the system state and the DQN cost, while transitions are stored in replay memory and sampled to update the Q-network. Finally, the trained DNN and DQN are returned for dynamic (online) sensing and beam~prediction.
\begin{algorithm}[t]
\caption{\small Proposed Joint Sensing and Beam Prediction Scheme}
\label{alg:dnn_dqn_training}
\small
\tcc{Initialization}
Set system parameters $V$, sensing limit, and initialize DNN and DQN parameters\;

\tcc{Step (1): Train DNN for beam prediction}
\For{each epoch}{
    Sample a mini-batch from the dataset\;
    Compute predictions and cross-entropy loss\;
    Update DNN parameters via backpropagation\;
}

\tcc{Step (2): Train DQN for sensing decisions}
\For{each episode}{
    Initialize environment and state $S(0)$\;
    \For{each time step $t=1,2,\dots,T$}{
        Choose action $x(t)$ using the $\epsilon$-greedy policy\;
        Execute $x(t)$: update $Q(t+1)$ by \eqref{eq_virtualQueue}, $\mathcal{X}(t)$ by \eqref{eq_ml_input}, and age $\theta(t+1)$ by \eqref{eq_agedynamic}, then obtain $S(t+1)$\;
        Use $\mathcal{X}(t)$ and the trained DNN (Step 1) to compute the cross-entropy loss in \eqref{eq_entropyloss}\;
        Compute the cost function in \eqref{eq_dqn_cost}\;
        Store transition $(S(t), x(t), C(t), S(t+1))$ in replay memory\;
        Sample a mini-batch from replay memory and update DQN parameters via Q-learning\;
        Set $S(t) \leftarrow S(t+1)$\;
    }
}

\tcc{Output}
Return the trained DNN and DQN for real-time inference~(Fig.~\ref{fig_proposed_method})\;
\end{algorithm}

\vspace{-0.25cm}
\section{Numerical Results and Discussions} \label{sec_numres}
This section presents the simulation results to evaluate the performance of the proposed method (see Fig.~\ref{fig_proposed_method}).\footnote{The source code is available at \url{https://github.com/AZakeri94/DQN_ML_Dynamic_Sensing-Communication.git}} For comparison, we consider two benchmarks. The first is \textit{randomized sensing}, where the sensing decision $x(t)$ is chosen randomly while still satisfying the constraint in \eqref{eq_cons_sensing}. The second is the \textit{without age} case, where the age of samples is excluded from the DQN state $S(t)$.  Performance is evaluated using Top-$k$ accuracy, defined as the probability that the optimal beam lies within the top-$k$ predicted beams.

For the sensing modality, we use position data from the DeepSense dataset (Scenario~5) \cite{Ahmed_deepsense}. The beam prediction DNN module follows the pretrained model in \cite{Ahmed_vision}, implemented as a two-layered multilayer perceptron. The input dimension matches the feature size, with each hidden layer containing 1024 neurons and ReLU activation at each layer. Training is performed using the Adam optimizer with a learning rate of $0.01$ and a batch size of 32.

For the DQN module, we employ a three-layer fully connected network with 128 neurons in each hidden layer. The input dimension matches the DQN state size, and the output dimension corresponds to the action space. Training is performed with a discount factor of $0.99999$, a learning rate of $0.001$ using the Adam optimizer, a batch size of 64, and a replay memory of 50{,}000. The model is trained for 300 epochs, each consisting of 400 iterations.

\begin{figure}[t]
  \centering
  \includegraphics[width=7cm]{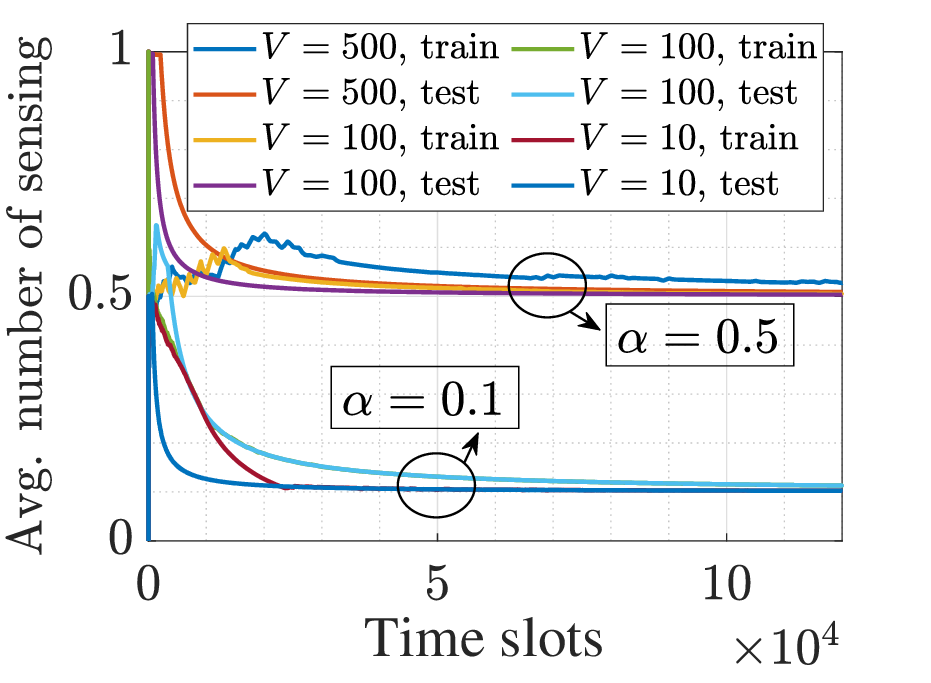}\vspace{-0.25cm}
  \caption{Satisfaction of the average constraint~\eqref{eq_cons_sensing} by the proposed DQN method for different $V$ and $\alpha$}
  \label{fig_cons_satis}
\end{figure}

We first demonstrate in Fig.~\ref{fig_cons_satis} that the proposed DQN-based sensing module satisfies the average constraint in \eqref{eq_cons_sensing} for different values of the normalized sensing budget ${\alpha \triangleq \dfrac{c^{\mathsf{max}}}{c}}$ with $\alpha \in [0,1]$. The results confirm that the cost term introduced in \eqref{eq_entropyloss} effectively enforces the average constraint. It is further observed that larger values of $V$ yield slower convergence to the constraint limit~$\alpha$.

\begin{figure}[t]
  \centering
  \includegraphics[width=7cm]{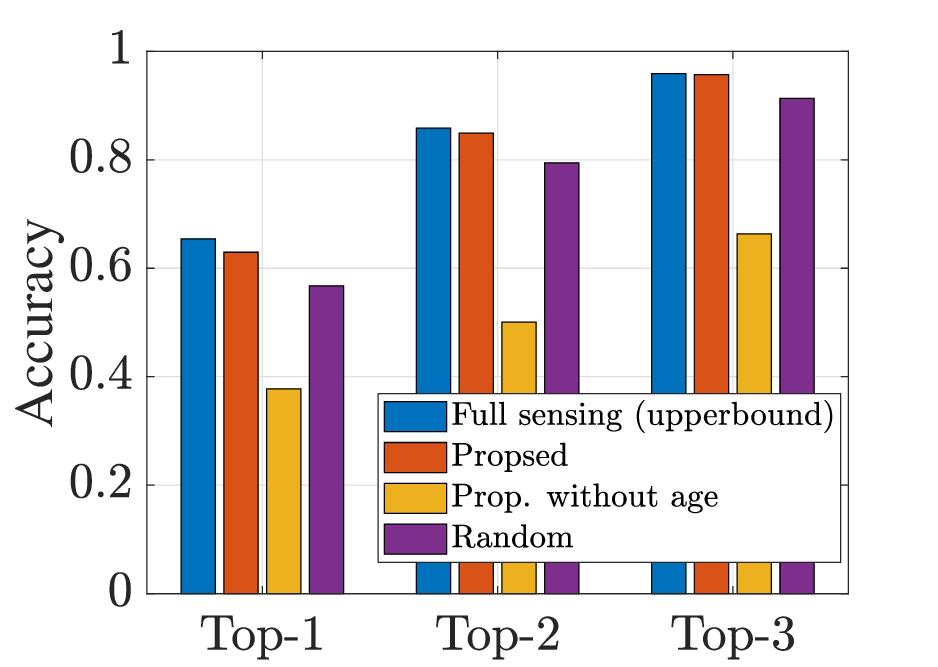}\vspace{-0.25cm}
  \caption{Accuracy comparison between different methods for the normalized sensing budget $\alpha=0.5$ and $V=100$}
  \label{fig_topk_accu_}
\end{figure}

Fig.~\ref{fig_topk_accu_} shows the Top-1, Top-2, and Top-3 beam prediction accuracies for different methods, including the full-sensing case where sensing is performed at every time slot. Two main observations arise: (i) reducing the sensing frequency by 50\% leads to almost no loss in beam prediction accuracy, and (ii) incorporating sensing data age into the DQN state yields a notable performance gain. The latter underscores the importance of the age of information \cite{Roy_2012} in improving remote inference for machine learning–driven communication tasks.

\begin{figure}[t]
  \centering
  \includegraphics[width=7cm]{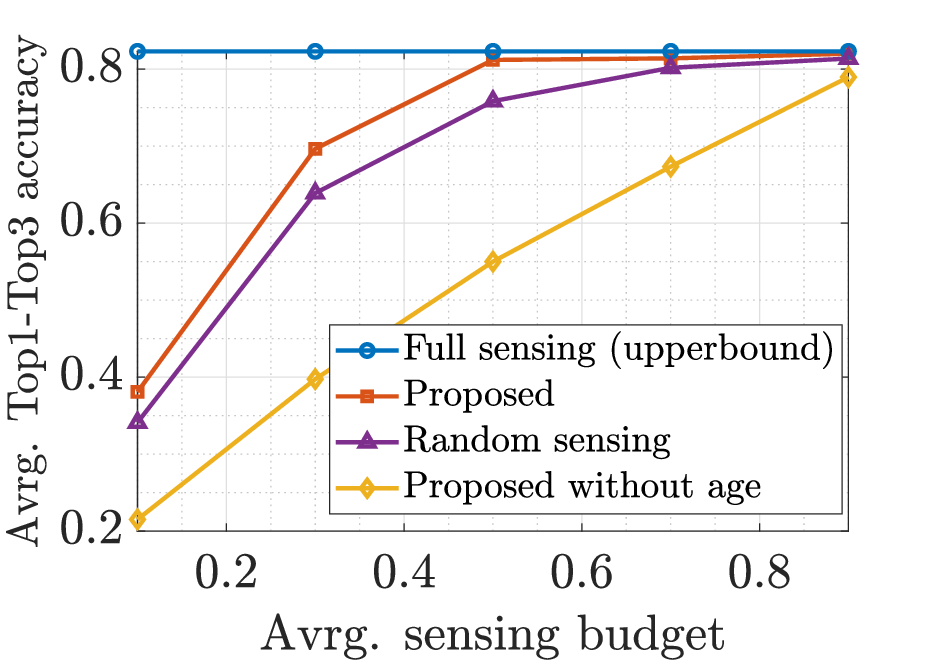}\vspace{-0.25cm}
  \caption{Average Top1-Top3 accuracy vs. the sensing budget.}
  \label{fig_avrgaccu_vs_alpha}
\end{figure}

Fig.~\ref{fig_avrgaccu_vs_alpha} compares the average accuracy, defined as $(\text{Top-1}+\text{Top-2}+\text{Top-3})/3$, of different sensing methods versus the sensing budget. The results also highlight the benefit of incorporating the age of the last sensing data into the DQN state, showing consistent gains over randomized sensing.~Moreover, our method converges to the full-sensing performance when $\alpha \geq 0.5$, corresponding to a 50\% reduction in sensing cost.
\section{Conclusions}\label{se_concl}
We proposed a dynamic sensing-aided communication framework that combines Lyapunov optimization with deep reinforcement learning for beam prediction under an average sensing cost budget. The framework integrates a DNN for beam prediction with a DQN for sensing decisions, where the DQN state incorporates the virtual queue, the DNN input, and the age of the most recent sensing data. A customized cost function was designed by combining cross-entropy with the Lyapunov drift. Evaluations on the DeepSense Scenario~5 dataset showed that the proposed method reduces sensing cost by nearly~50\% without sacrificing beam prediction accuracy, and that incorporating sensing data age yields significant gains. Future work will extend the approach to general multimodal sensing scenarios to further assess its effectiveness.

\vfill\pagebreak
\section{Acknowledgment}
This work was supported by the Research Council of Finland through the 6G Flagship Program (Grant No. 369116),  
project DIRECTION (Grant No. 354901), project DYNAMICS (Grant No. 367702), and project S6GRAN (Grant No. 370561);  
supported in part by CHIST-ERA through the project PASSIONATE (Grant No. 359817);  
and in part by the HORIZON-JU-SNS-2023 project INSTINCT.

\bibliographystyle{ieeetr}
\bibliography{Bio/conf_short,
Bio/IEEEabrv,
Bio/Bibliography, Bio/ML_Bio, Bio/multimodalsensing_Bio }

\end{document}